\documentclass[prd,aps,twocolumn,floats]{revtex4}
\usepackage{amssymb,graphicx,epsfig}
\newcommand{\p}{\partial}
\newcommand{\be}{\begin{equation}}
\newcommand{\ee}{\end{equation}}
\newcommand{\bea}{\begin{eqnarray}}
\newcommand{\eea}{\end{eqnarray}}
\newcommand{\beay}{\begin{eqnarray*}}
\newcommand{\eeay}{\end{eqnarray*}}

\newcommand{\SBP}{SBP}

\begin{document}

\title{Novel finite-differencing techniques for numerical relativity:\\ 
application to black hole excision}
\author{Gioel Calabrese$^1$, Luis Lehner$^1$, David Neilsen$^1$, Jorge
Pullin$^1$, Oscar Reula$^2$, Olivier Sarbach$^1$, and Manuel Tiglio$^1$}
\affiliation{$1$ Department of Physics and Astronomy, Louisiana State
University, 202 Nicholson Hall, Baton Rouge, LA 70803-4001\\
$2$ FaMAF, Universidad Nacional de Cordoba, Cordoba, Argentina 5000}

\begin{abstract}
We use rigorous techniques from numerical analysis of hyperbolic
equations in bounded domains to construct stable finite-difference
schemes for Numerical Relativity, in particular for their use in black
hole excision.  As an application, we present 3D simulations of a
scalar field propagating in a Schwarzschild black hole background.
\end{abstract}

\maketitle The numerical implementation of Einstein's equations
represents a daunting task.  The involved nature of the equations
themselves, and a number of technical difficulties (related to the
necessarily finite computational domain, the limitations in relative
computational power and the need to deal with singularities,
constraints and gauge freedom) imply a significant challenge.

Numerical solutions of Einstein's equations involve solving a nonlinear
set of partial differential equations on a bounded domain, and formally
constitute an initial-boundary-value problem (IBVP).  
Constructing stable and long term well behaved numerical 
approximations for such systems with boundaries is highly
non-trivial. Here the term numerical stability is used in the
sense that is equivalent, through Lax's theorem, to convergence (that
is, the property that the numerical solution will approach the
continuum one when resolution is increased). 
Delicate complications arise due to corner and edges at outer (and
possible inner) boundaries, all of which introduce subtleties 
for a stable implementation.

Recently, however, several sophisticated tools of rigorous numerical
analysis have been developed for systematically constructing stable
numerical schemes for IBVP's.  They employ a discrete form of
well-posedness and thus are stable by construction, at least for
linear systems. At this time, these tools have essentially not been
used by the numerical relativity community. The purpose of this paper
is to outline their use, in particular for black hole excision.

An IBVP consists of three ingredients: a partial differential
equation, initial and boundary data. It is well-posed if a 
solution exists, is unique, and depends continuously on the initial 
and boundary data. It is well known
that stable finite difference schemes approximating an IBVP can
only be constructed for well-posed systems.
While problems in general relativity are typically nonlinear we consider
here the linear IBVP, as stability in the linearized case
is a necessary condition for stability in the full nonlinear system 
and these methods may also be applied to nonlinear equations.  
Consider the linear IBVP on a domain 
$[0,\infty) \times \Omega$
\begin{eqnarray}
   \partial _t u &=& A(t,\vec{x}) ^i \partial _i u 
                             + B(t,\vec{x})u  \label{evol_eq},\\
    u(0,\vec{x}) &=& f(\vec{x}), \\
     w_+(t,\vec{x}) &=& S
w_-(t,\vec{x}) + g(t,\vec{x}) \;,  \vec{x} \in \partial \Omega ,\label{eq:bc}
\end{eqnarray}
where $u$ is a vector-valued function, $w_+$ and $w_-$ are incoming
and outgoing modes, and $S$ is sufficiently small (maximally
dissipative boundary conditions~\cite{maxdissbc}).
The system is assumed to be symmetric hyperbolic, i.e., 
there exists a symmetric positive definite
matrix $H(t,\vec{x})$, the symmetrizer, satisfying $HA^i = (HA^i)^T$.  
The usual proof of well-posedness proceeds by
defining an ``energy'', ${\cal E} = \int_{\Omega} u^TH u \,d^3x\,$,
and by showing that ${\cal E}$ can be bounded as a
function of the initial and boundary data. 
Analogously, a way to
construct stable numerical schemes is to design them such that
 energy estimates hold at the discrete level \cite{gko}. This is the procedure we
shall follow in this paper.

Our construction involves four steps. (1) We construct discrete
derivative operators and a scalar product so that a semi-discrete
energy estimate holds.  The spatial derivatives are then discretized
using these operators, resulting in a semi-discrete system of ordinary
differential equations.  (2) We impose boundary conditions in a way
that does not spoil the previous semi-discrete energy estimate. In
particular, we account for effects of boundary edges and
vertices. Steps (1) and (2) guarantee numerical stability of the
semi-discrete system. (3) We may add artificial dissipation and/or
arrange the spatial discretization to achieve optimal energy bounds.
(4) Finally, an appropriate time integrator is chosen so that {\it
fully discrete stability} holds.

{\em Step (1):} The equations are discretized on a domain $\Omega$
with an inner boundary to accommodate for black hole excision.  While
several grid geometries are possible, we concentrate on the simplest
case of a cubical domain from which an inner, smaller cube has been
removed (both cubes aligned with the grid).  We introduce the grid
points $\vec{r}_{ijk} = (i\Delta x, j\Delta y, k \Delta z) \in \Omega$
and assume that some of these points lie on the boundary $\p\Omega$.
A scalar product, $\Sigma$, between any two real vector-valued grid
functions $u$ and $v$, is defined as \be (u,v)_{\Sigma } = \Delta x
\Delta y \Delta z \sum^N_{i,j,k=0}\sigma_{ijk} u^T_{ijk} v_{ijk},
\label{sp} \ee where $\sigma_{ijk}$ are defined below and
in the sum $\vec{r}_{ijk}\in \Omega$.  A semi-discrete energy is then defined by
$E = (u,Hu)_{\Sigma }$.

A key ingredient for deriving the continuum energy estimate
is integration by parts. Similarly, in
order to get a semi-discrete energy estimate, one must 
construct the difference
operators, $D$, and $\Sigma$ in such a way that the discrete version of
integration by parts, called {\em summation by parts} (SBP),
holds. In one dimension, \SBP\  is expressed by 
$(u,Dv)_{\Sigma} + (Du,v)_{\Sigma} = u_Nv_N - u_0v_0$,
and this definition can be generalized to higher dimensions and more
complicated domains.

A detailed calculation shows that SBP for our chosen $\Omega$ holds by defining
(similar definitions for $y,z$ directions)
$$ D^{(x)} = \left\{
\begin{array}{l@{\quad}l}
D^{(x)}_{\pm} u_{ijk} & \mbox{at $x=\mbox{const.}$ faces,}\\
& \mbox{$x=\mbox{const.}$ OB edges}\\
& \mbox{and OB vertices}\\
\left(\frac{1}{3}D^{(x)}_{\pm} + \frac{2}{3}D^{(x)}_0\right) u_{ijk} &
\mbox{at $x=\mbox{const.}$ IB edges} \\
\left(\frac{1}{7}D^{(x)}_{\pm} + \frac{6}{7}D^{(x)}_0\right) u_{ijk} &
\mbox{at IB vertices}\\
D^{(x)}_0 u_{ijk} & \mbox{everywhere else}
\end{array}
\right.
$$
where IB and OB  stand for inner and outer boundary
respectively, $D^{(x)}_+ u_{ijk} = (u_{i+1\,jk}-u_{ijk})/\Delta x$, $D^{(x)}_-
u_{ijk} = (u_{ijk}-u_{i-1\,jk})/\Delta x$ and $D^{(x)}_0 u_{ijk} =
(u_{i+1\,jk}-u_{i-1\,jk})/(2\Delta x)$.  The $\pm$ 
signs indicate whether non-excised points are to the right or left, respectively, 
of the boundary point.  
The weights $\sigma_{ijk}$ in (\ref{sp}) are
 defined  to be  $1$ in the interior, $1/2$ at the faces of IB and OB,
 $1/4$ at the edges of OB,
$3/4$ at the edges of IB, $1/8$ at the vertices of OB, 
$7/8$ at the vertices of IB, and zero in the 
excised region. 
This difference operator is second-order accurate at the interior and
first-order at boundaries, yielding in principle, a scheme with overall 
second-order accuracy\cite{gus}.

{\em Step (2):}
In contrast to the continuum problem, an examination
of boundary terms left after \SBP\ in the energy estimate indicates that edge
and vertex boundary points make a  finite contribution to $E$ at fixed
resolution.  These contributions show precisely how to impose boundary
conditions at these points. They have to be imposed on incoming
characteristic modes defined by  some ``effective'' normal vectors
$n$.  
For instance, on a {\em uniform} grid ($\Delta x = \Delta y =
\Delta z$), at an edge  one has
$n= (n_a+n_b)/\sqrt{2}$, and at a
vertex $n= (n_a+n_b+n_c)/\sqrt{3}$ (where $n_a$, $n_b$ and $n_c$ are unit
vectors of the intersecting faces). Boundary conditions along these effective
directions have to be imposed without destroying the semi-discrete
energy estimate. Following Olsson \cite{olsson}, we do so 
by projecting the right hand side (RHS) of the evolution
equations at boundary points onto the 
subspace of the grid functions
that satisfy the discretized boundary conditions. The projection is
by construction self-adjoint with respect to $\Sigma$, which 
ensures that the solution of the projected semi-discrete system 
will satisfy the boundary conditions without compromising numerical 
stability.

{\em Step (3):} 
At a fixed resolution,
even convergent codes may generate significant error
growth as time progresses, and it is often
desirable to minimize this growth.  This may be
achieved by adding artificial dissipation, by rearranging the
semi-discrete 
equations in a specific manner to obtain optimal estimates, 
or by doing both. We briefly summarize these two procedures.

In numerical simulations a dissipative term,
$(Q^{(x)}_d + Q^{(y)}_d + Q^{(z)}_d) u$, 
is sometimes added to 
the RHS of Eq.~(\ref{evol_eq}),
which damps high frequency modes but does not
affect the accuracy of the scheme. Requiring
 $(u,Q^{(x)}_d u)_{\Sigma } \le 0$
 ensures that the semi-discrete energy estimate still holds.
We have modified the
Kreiss--Oliger dissipation operator
for our black hole excision geometry and $\Sigma$. 
This modified operator has the standard form on the grid interior
(shown here for the $x$ direction)
\beay
Q^{(x)}_d u_{ijk} &=& -\epsilon \Delta x^3 (D^{(x)}_+D^{(x)}_-)^2
u_{ijk},
\eeay
(where $\epsilon$ is a parameter satisfying $\epsilon \geq 0$.)
However, it is modified near boundaries to satisfy  $(u,Q^{(x)}_d u)_{\Sigma } \le 0$.
Let $(i_0,j,k)$ label a point on a $x=\mbox{constant}$ boundary with neighbors 
$(i_0-1,j,k)$ and $(i_0+1,j,k)$.  Near this boundary point $Q^{(x)}_d$ becomes
\beay 
&&Q^{(x)}_du_{i_0-1} = -\epsilon \Delta x
(D^{(x)2}_- - 2D^{(x)}_+D^{(x)}_-)u_{i_0-1},\\
&&Q^{(x)}_du_{i_0} = -\frac{\epsilon \Delta x}{\sigma_{i_0}} 
(\sigma _{i_0-1} D^{(x)2}_- + \sigma _{i_0+1} D^{(x)2}_+)u_{i_0},\\
&&Q^{(x)}_du_{i_0+1} = -\epsilon \Delta x (D^{(x)2}_+ - 
2D^{(x)}_+D^{(x)}_-)u_{i_0+1},
\eeay
where the indices $jk$ are suppressed for clarity, and
$Q^{(x)}_d$ is set to zero for points outside the domain.

A second method for controlling unnecessary growth is by rearranging
the discretized equations so the optimal estimates for the continuum
hold at the semi-discrete level. For instance, if at the continuum the system exhibits
energy conservation, it is important to preserve it at the
semi-discrete level. For example, assuming that
Eq.~(\ref{evol_eq}) has time-independent coefficients and that
$\p_i(HA^i) = HB + (HB)^T$, ${\cal E}$ is conserved if appropriate
boundary conditions are given.  The solution of $\p_t u = A^i(x) D_i u
+ B(x) u$ may yield growth in $E$, due 
to the lack of a Leibniz rule at the discrete level.
However, one can show that rearranging the semi-discrete RHS of
(\ref{evol_eq}) as
$$
\frac{1}{2}A^iD_i u + \frac{1}{2} H^{-1} D_i (HA^i u)
+\left(B - \frac{1}{2}H^{-1}\p_i(HA^i)\right)u,
$$
gives a non-increasing $E$, and therefore the discrete solution will
not grow. This idea is closely related to the concept of {\it strict
  stability} \cite{olsson}.

{\em Step (4):}
Finally, stability of the fully discrete system follows if it
is integrated with a scheme that satisfies the (necessary and
sufficient) {\it local stability} condition or the (sufficient) preservation of
energy estimate's condition  (e.g., third or fourth order Runge--Kutta \cite{rk}).

{\em An example:} Wave propagation on a black hole spacetime
$g_{\mu\nu}$ presents some of the same challenges as the full Einstein
equations (e.g. constraint preservation, boundary conditions, excision
of the singularity). Thus it is an ideal test bed to demonstrate the
techniques described in this paper. We write the wave equation in
first-order form,
\begin{equation}
\nabla_\mu\Phi = V_\mu\; ,\quad
\nabla^\mu V_\mu = 0\;, \quad
\nabla_\mu V_\nu = \nabla_\nu V_\mu\; ,
\end{equation}
with $\nabla_{\mu}$ the covariant derivative.
To split these equations into evolution equations and constraints,
we specify a future-directed time-like vector  $u^\mu$,
and contract the first and the last equation with it.
The evolution equations are
\begin{equation}
\pounds_u\Phi = u^\mu V_\mu \equiv \Pi , \quad
\nabla^\mu V_\mu = 0, \quad
\pounds_u V_\mu = \nabla_\mu\Pi,
\label{Eq:WE}
\end{equation}
where $\pounds_u$ is the Lie derivative with respect to $u^\mu$.
This system is symmetric hyperbolic.
Due to the evolution equations, 
the constraints $C_\mu \equiv V_\mu -
\partial_\mu\Phi=0$, $C_{\mu\nu} \equiv \nabla_\mu V_\nu - \nabla_\nu
V_\mu=0$ are Lie-dragged by $u^\mu$, i.e., $\pounds_u C_\mu = 0$, $\pounds_u
C_{\mu\nu} = 0$.

\begin{figure}[t]
\epsfig{file=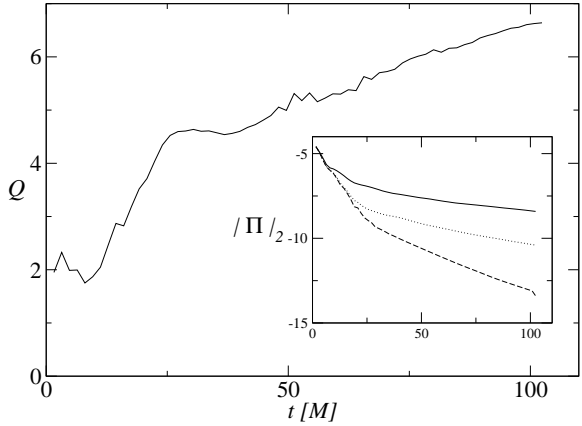,height=2.2in,angle=0}
\caption{
The self-convergence factor, $Q$, as a function of time, where
$Q \equiv \log_2(||f_{\Delta x} - f_{2\Delta x}||/||f_{2\Delta x} - f_{4\Delta x}||)$,
and $f_{\Delta x}$, $f_{2\Delta x}$, and $f_{4\Delta x}$ represent numerical solutions
calculated on uniform grids with corresponding gridspacing.
The inset shows the $L_2$ norm of the
variable $\Pi$ vs. time for the three resolutions.  The solution is
calculated on the domain $x^i\in[1.5M,5.5M]$, using $41^3$, $81^3$ and $161^3$
grid points, dissipation parameter $\epsilon=0$, and Courant factor $\lambda=0.8$. 
Since we are outside the hole,
we chose $b^i=0$, giving a symmetric hyperbolic
formulation. $\Pi$ has non-zero initial data of compact support, and all
other fields are initially set to zero. The solution is essentially second order
convergent until it decreases several orders of magnitude. After that
the self-convergence factor grows and the evolution is followed
roughly until the solution for the finest resolution reaches
truncation error (about $25$ crossing times).}
\end{figure}

Although the discussion below applies to any background geometry, for
definiteness we specialize to the Schwarzschild background. This
spacetime possesses a time-like Killing field, $k^\mu =
(\partial_t)^\mu$, and the future directed unit normal $n^{\mu}$ to
the $t = \mbox{const.}$ slices is given by $(k^\mu -
\beta^\mu)/\alpha$. We denote with $\alpha, \beta^i$ and $h_{ij}$ the
lapse, shift and three-metric, respectively. A natural choice for the
vector field $u^\mu$ is $n^\mu$, however this requires care specifying
boundary data compatible with the constraints.  For a particular
coupling between the in- and out-going modes, we have found maximally
dissipative boundary conditions compatible with the
constraints. Unfortunately, they imply reflective boundary conditions
in the sense that (for homogeneous boundary data and in the absence of
inner boundaries) the physical energy of the scalar field $\Phi$ is
exactly conserved. Choosing a different formulation with $u^\mu =
k^\mu$ allows us to impose radiative boundary conditions, in the sense
that the physical energy decreases when the wave reaches the boundary.
In this formulation the constraint variables $C_\mu$ and $C_{\mu\nu}$
propagate tangentially to the boundaries, and thus the constraints are
automatically satisfied when satisfied initially. Moreover, this
formulation has the advantage that its symmetrizer agrees with the
physical energy. Unfortunately, as the Killing field $k^\mu$ becomes
space-like inside the black hole, the symmetrizer is not positive
definite in this region.

\begin{figure}[t]
\epsfig{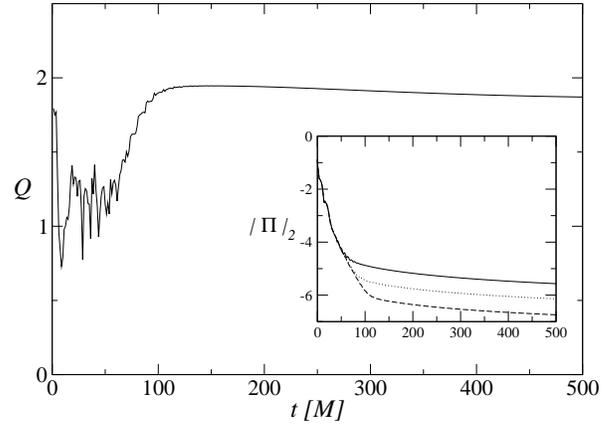}
\caption{This figure shows the self-convergence factor defined in
Fig.~1, Q.  A non-trivial $b^i$ smoothly interpolating is used so that
the system is everywhere symmetric hyperbolic.  The domain is
$x,y,z\in[-4M,4M]$, with an excised region $x,y,z\in[-0.375M,
0.375M]$.  The Courant factor and dissipation are $\lambda=1$,
$\epsilon=0.02$, respectively, and runs with $65^3$, $129^3$ and
$257^3$ grid points are used to calculate $Q$.  The initial part of
the plot shows lower than second order convergence. However,
experiments in one dimension show a similar effect for comparable grid
spacings; though the convergence factor asymptotically approaches two
as resolutions are increased.}
\label{fig2}
\end{figure}

We combine the advantages of both formulations by choosing $u^\mu =
(k^\mu - b^\mu)/\alpha$, where $b^\mu$ is a smooth vector field which
is tangential to the $t=\mbox{const.}$ slices, vanishes in a
neighborhood of the outer boundary, and agrees with the shift vector
$\beta^i$ as one approaches the event horizon from the outside region.
This new interpolating formulation is manifestly symmetric hyperbolic 
also inside the black hole; and at the outer boundary radiative boundary conditions
can be given. Equations (\ref{Eq:WE}) yield
\begin{eqnarray}
\partial_t \Pi &=& b^i\partial_i\Pi + \frac{\Delta^i}{\alpha} \partial_i(\alpha\Pi)
  + \frac{1}{\sqrt{h}}\partial_i (\sqrt{h}\Delta^i\Pi) \nonumber\\
 &+& \frac{1}{\sqrt{h}} \partial_i\left( \alpha\sqrt{h} H^{ij} V_j \right)
  - \frac{1}{\sqrt{h}}\partial_j\left( \sqrt{h}\,\frac{b^j}{\alpha} \right) \Delta^i V_i \nonumber\\
 &+& \frac{1}{\alpha}\left(\beta^j\partial_j b^i - b^j\partial_j\beta^i\right) V_i
  + \frac{1}{\sqrt{h}} \partial_i (\sqrt{h} b^i)\Pi\label{Eq:Int2}\\
\partial_t V_i &=& \partial_i(\alpha\Pi) + b^j\partial_j V_i
 + V_j\partial_i b^j \label{Eq:Int3}
\end{eqnarray}
where $\Delta^i = \beta^i - b^i$,
$H^{ij} = h^{ij} - \Delta^i\Delta^j/\alpha^2$, and  $h^{ij}$ denotes
the inverse three metric. This system is symmetric hyperbolic with
respect to the energy ${\cal E} = \frac{1}{2} \int_\Omega \alpha\left(
\Pi^2 + H^{ij} V_i V_j \right) \sqrt{h} d^3 x$.  The change of ${\cal
E}$ under the flow generated by (\ref{Eq:Int2},\ref{Eq:Int3}) can be
bounded as
\begin{equation}
\frac{d}{dt} {\cal E} \leq \int\limits_{OB}
\frac{1}{4}\left( w_+^2 - w_-^2 \right) \sqrt{h} d^2 x + \int_\Omega F d^3 x,
\label{eq:sfestimate}
\end{equation}
where $w_{\pm} = \mu_{\pm} \Pi + \alpha H^{ij} n_i V_j$ are the
characteristic variables which have nonzero speeds, $\mu_{\pm} =
\pm\alpha + \beta^i n_i$, and $n_i$ is the normal to the outer boundary,
normalized such that $h^{ij} n_i n_j = 1$.  The expression
$F$ in Eq.~(\ref{eq:sfestimate}) vanishes if $b^i = 0$.  Then
setting $w_+$ to zero guarantees that ${\cal E}$ will decrease when
the wave reaches the boundary.
In the previous estimate we have assumed that the excised cube is
sufficiently small (side length smaller than $4\sqrt{3}M/9$)
so that the inner boundary is purely outflow.

Equations ~(\ref{Eq:Int2})--(\ref{Eq:Int3}) have been written such that
the replacement $\p_i$ by $D_i$ (only on the dynamical variables)
automatically leads
to a semi-discrete system with  energy conservation in the case $b^i=0$.
The semidiscrete solution will not grow, even at fixed resolution. 

The figures show the results of the numerical implementation of the
techniques for the scalar wave example we discussed; we have chosen
Kerr-Schild coordinates (we have checked that the case of
Painlev\'e-Gullstrand coordinates behaves similarly and so does the
Maxwell field on these backgrounds) . Figure~1 shows evolutions with
the computational domain outside the event horizon; fig.  2 shows
results of a run with singularity excision. Movies can be see in
\cite{movie}.

In this article we have shown that the use of rigorous numerical
analysis techniques can be used in practice to obtain stable schemes
for numerical relativity. This maps out a precise road for robust
implementations, reducing the need for expensive trial and
error standard methods. The successful application of these techniques
to the simulation of fields  propagating on  black hole backgrounds, which
requires dealing with singularity excision, corners and edges,
illustrates the potential of the approach. This opens the door for
applying the same techniques to the full equations of general
relativity.

{\it Acknowledgments:} We thank E. Tadmor, P. Olsson, and
H. O. Kreiss for helpful discussions. This work was supported in part
by NSF grant PHY9800973, the Horace Hearne Jr. Institute for
Theoretical Physics, and Fundaci\'on Antorchas.  Computations were
done at LSU's Center for Applied Information Technology and Learning
and parallelized with the CACTUS toolkit \cite{cactus}. GC, LL, OR, OS and MT thank
the Caltech Visitors Program in numerical relativity for hospitality.


\end{document}